\documentclass[12pt]{article}
\usepackage{amsmath,amsfonts,amssymb,bbm}
\usepackage{tensor}
\usepackage[T1]{fontenc}
\usepackage[utf8]{inputenc}
\usepackage{lmodern}
\usepackage{authblk}
\usepackage[disable]{todonotes}
\usepackage{color}
\usepackage{multicol}
\usepackage{cite}
\usepackage[margin=3cm]{geometry}
\usepackage{hyperref}

\newcommand\cM{{\cal M}}

\newcommand{\bz}{\bar{z}}

\newcommand{\p}{\partial}
\newcommand{\pb}{\p_{\bz}}
\newcommand{\be}{\begin{equation}}
\newcommand{\ee}{\end{equation}}
\def\bea{\begin{eqnarray}}
\def\eea{\end{eqnarray}}
\def\nn{\nonumber}
\def\bm{\bar{m}}
\newcommand{\half}{\frac12}
\def\n{\nabla}
\def\bP{\bar P}

\def\pu{{}}
\def\pj{{}}
\allowdisplaybreaks[1]
\newcommand*\xbar[1]{%
  \hbox{%
    \vbox{%
      \hrule height 0.5pt 
      \kern0.3ex
      \hbox{%
        \kern-0.0em
        \ensuremath{#1}%
        \kern-0.0em
      }%
    }%
  }%
}

\newcommand\email[1]{\thanks{\href{mailto:#1}{\nolinkurl{#1}}}}

\author[a,c]{Pujian Mao\email{pjmao@tju.edu.cn}\,}
\author[b,d]{Xiaoning Wu\email{wuxn@amss.ac.cn}\,}

\affil[a]{\,Center for Joint Quantum Studies and Department of Physics, School of Science, Tianjin University, 135 Yaguan Road, Tianjin 300350, P. R. China}

\affil[b]{\,Institute of Mathematics, Academy of Mathematics and System Science, Chinese Academy of Sciences, Beijing 100190, China}

\affil[c]{\,Institute of High Energy Physics and Theoretical Physics Center for Science Facilities, Chinese Academy of Sciences, 19B Yuquan Road, Beijing 100049, P. R. China}

\affil[d]{\,Hua Loo-Keng Key Laboratory of Mathematics, Chinese Academy of Sciences, Beijing 100190, China}

\title{\bf More on gravitational memory\\}
\date{}

\begin{document}

\maketitle
\thispagestyle{empty}

\begin{abstract}
Two novel results for \pj{the} gravitational memory effect are presented in this paper. We first extend the formula \pj{for} the memory effect to solutions with arbitrary two surface boundary topology. The memory effect for \pj{the} Robinson-Trautman solution is obtained in its standard form. Then we propose a new observational effect for the spin memory. It is a time delay of time-like free falling observers.
\end{abstract}



\section{Introduction}

\pj{The} gravitational memory effect was first reported by Zel'dovich and Polnarev \cite{memory} in linearized gravity and further investigated by Christodoulou in full Einstein gravity \cite{Christodoulou:1991cr} (see also \cite{Braginsky:1986ia,1987Natur,Wiseman:1991ss,Thorne:1992sdb,Frauendiener} for further development and \cite{Lasky:2016knh,Nichols:2017rqr,Yang:2018ceq} for the realization with gravitational wave detectors). The memory effect is a relative displacement of nearby observers. It is therefore called \pj{the} displacement memory effect. Recently, memory effect\pj{s have obtained} renewed interest from \pj{a} purely theoretical point of view. A fundamental connection between \pj{the} displacement gravitational memory effect and Weinberg's soft graviton theorem \cite{Weinberg:1965nx} was discovered by Strominger and Zhiboedov \cite{Strominger:2014pwa}\footnote{See also the analogue in gauge theory \cite{Bieri:2013hqa,Pasterski:2015zua,Susskind:2015hpa,Mao:2017axa,Mao:2017wvx,Pate:2017vwa,Ball:2018prg,Afshar:2018sbq}}. The gravitational memory formula and the Fourier transformation of Weinberg's soft graviton formula are mathematically equivalent.

\pj{A} recent investigation on soft graviton theorems \cite{Cachazo:2014fwa} shows that the universal property goes beyond Weinberg's pole formula and contains next-to-leading orders in the low-energy expansion. Inspired by the sub-leading soft graviton theorem, a new gravitational memory was proposed in \cite{Pasterski:2015tva}. This new gravitational memory effect is suggested to be a relative time delay between different orbiting light rays induced by radiative angular momentum flux. Accordingly it is called the spin memory.

The standard treatment of the memory effect \cite{Christodoulou:1991cr,Frauendiener} is based on a special choice of the topology of null infinity which is $S^2\times \mathbb{R}$. From the geometrical point of view, null infinity is not part of space-time but can be added to it by conformal compactification \cite{Penrose:1962ij,Penrose:1965am}. Hence, the topology of any asymptotically flat space-time can be always set to be the standard $S^2\times \mathbb{R}$ via changing the conformal factor during the compactification. However, asymptotically flat solutions may not be in their simplest form with the boundary topology being unit 2 sphere but \pj{an} arbitrary 2 surface. Relating those solutions to the standard boundary topology will lose the simplicity both geometrically and algebraically. For instance, the well-known Robinson-Trautman waves \cite{Robinson:1960zzb} will not be truncated in the $\frac1r$ expansion with a unit 2 sphere boundary topology. It is extremely difficult, if not impossible, to see the geometrical property of this simple but very important exact solution with gravitational radiation. Therefore, it is definitely meaningful to extend the formula of the memory effect to the case of arbitrary 2 surface boundary topology. This is precisely what we will show in following pages.

Another purpose of this paper is to provide a new observational effect of the spin memory. In \cite{Pasterski:2015tva}, it is proposed that light rays orbiting in different directions acquire a relative delay which will induce a shift in the interference fringe. Alternatively, we propose to examine the spin memory effect by time-like free falling observers who are very close to null infinity. We find that \pj{a} free falling observer, which is initially static, is forced to orbit by the gravitational radiation (see also \cite{Grishchuk:1989qa,Podolsky:2002sa,Podolsky:2010xh,Podolsky:2016mqg,Zhang:2017rno,Zhang:2017jma,Zhang:2018srn,Compere:2018ylh}). It receives \pj{a} time delay due to massive objects in the space-time, \textit{e.g} massive stars or black holes, and gravitational radiation (see also \cite{Flanagan:2018yzh}). The former is the well-known Shapiro time delay \cite{Shapiro:1964uw} (see \cite{Visser:1998ua,Visser:1999fe} for recent development) and was observationally verified almost 40 years ago \cite{Reasenberg:1979ey}, while the \pj{latter} is less stressed elsewhere. If a ring of freely falling observers who are initially static and synchronized can be set, the changes of the proper time of every observer on this ring will be different at later time. It can therefore memorize the waveform of the gravitational waves. Though \pj{a} stationary massive object can cause \pj{a} time delay \pj{for a} single observer, the change \pj{in} proper time of \pj{each} observer \pj{is the same}. Consequently, the ring of free falling observers will \pj{detect} two memory effects: the displacement memory that will squash and stretch the shape of the ring and the time delay that will cause \pj{a} difference in the proper time of nearby observers.

The plan of this paper is quite simple. In \pj{the} next section, we will derive the formula of the displacement memory with arbitrary 2 surface boundary topology. Section 3 will present the displacement memory of Robinson-Trautman waves as a precise example. In section 4, we compute the time delay formula of free falling time-like observers. Then we will prove that conjugate points on a time-like geodesic are very far from each other in the asymptotic region in section 5. Some comments will be given in the discussion section. There are also two appendices providing useful information for the main text.

\section{Displacement memory effect}

By setting the boundary topology to be $S^2\times \mathbb{R}$, the null basis vector $n$ in the standard Newman-Penrose formalism \cite{Newman:1961qr} is tangent to null geodesic\pj{s} with affine parameter $u$ on null infinity. The displacement memory effect in such case\pj{s} is controlled by the time integration of the asymptotic shear of $n$, \textit{i.e.} $\lambda^0$. This is equivalent to the change of the asymptotic shear of $l$, \textit{i.e.} $\sigma^0$, at early time $u_i$ and late time $u_f$ \cite{Frauendiener}. However, according to Newman-Unti \cite{Newman:1962cia} (see also Appendix \ref{NU solution}), the most general asymptotically flat solution is derived with the boundary topology being arbitrary 2 surface including the well-known Robinson-Trautman metrics \cite{Robinson:1960zzb}. Actually, Robinson-Trautman is very special as it allows \pj{a} shear-free null geodesic congruence, namely $\sigma=0$. From the standard formula, nothing will be memorized, but it is indeed an exact solution representing spherical radiation. Clearly, a general formula of displacement memory is of \pj{urgent} need for answering this type of question. This is what we will setup in this section.

We will first check the memory formula for the solutions with arbitrary 2 surface null boundary topology by \pj{mapping them to} the unit 2 sphere via Weyl transformations, \pj{as} is well studied very recently by Barnich and Troessaert \cite{Barnich:2016lyg}. In Appendix \ref{Weyl}, relevant results are presented and we will follow the convention of \cite{Barnich:2016lyg}. We use $(u,z,\bz)$ coordinates and unprimed quantities for \pj{the} unit 2 sphere case with boundary metric $ds^2=\frac{4}{(1+z\bz)^2}dzd\bz$ while $(u',z',\bz')$ coordinates\footnote{One should not confuse \pj{this} with the notation in Appendix \ref{NU solution} where $(u,z,\bz)$ is used for a general solution with arbitrary 2 surface boundary.} and primed quantities are for the solution in \pj{its} original form with arbitrary 2 surface boundary metric $ds^2=\frac{2}{P'\bP'}dz'd\bz'$. The two coordinates are connected in the following way\footnote{The full coordinate transformation is given in the form of \pj{an} asymptotic expansion, but only the leading terms are involved \pj{in} deriving the transformation law of \j{the} relevant fields.}
\be
z'=z,\;\; u'=\int^u_0 dv \;\frac{P_s}{\sqrt{P'\bP'}},\;\; P_s=\frac{1+ z \bz}{\sqrt{2}}.
\ee
Under such \pj{a} coordinate transformation, $\sigma'^0$ is transformed as
\be
\sigma_s^0=P_sP'^{-\frac32}\bP'^{\frac12}\sigma'^0 - \p_{\bz} \left(P_s \sqrt{P'\bP'} \p_{\bz} u' \right) + \sqrt{P'\bP'} \p_{\bz} u' \p_u \left(\sqrt{P'\bP'} \p_{\bz} u'\right).
\ee
$P',\;\bP'$ and $\sigma'^0$ are scalar fields, hence
\be
P'(u',z',\bz')=P(u,z,\bz),\;\bP'(u',z',\bz')=\bP(u,z,\bz),\;\;\sigma'^0(u',z',\bz')=\sigma^0(u,z,\bz).
\ee
We can just drop the prime
\be
\sigma_s^0=P_sP^{-\frac32}\bP^{\frac12}\sigma^0 - \p_{\bz} \left(P_s \sqrt{P\bP} \p_{\bz} u' \right) + \sqrt{P\bP} \p_{\bz} u' \p_u \left(\sqrt{P\bP} \p_{\bz} u'\right).
\ee
Since we have put the solution in $(u,z,\bz)$ coordinates with $S^2\times \mathbb{R}$ boundary topology, the standard displacement memory formula \pj{works}. It is just the change of $\sigma_s^0$ at early time $u_i$ and late time $u_f$ in this coordinates.

Alternatively, in $(u',z',\bz')$ coordinates, one can define
\be
{\sigma'}_s^0=P_s P'(u',z',\bz')^{-\frac32}\bP'(u',z',\bz')^{\frac12}\sigma'^0(u',z',\bz') + P_s \p_{\bz'}^2\left[\int^{u'}_0 dv \;\sqrt{P'\bP'}\right].
\ee
This is the Weyl invariant part of $\sigma'^0(u',z',\bz')$ \cite{Barnich:2016lyg}, namely it is unchanged as a function of their variables under Weyl transformation. This ${\sigma'}_s^0$ determines the memory effect in $(u',z',\bz')$ coordinates. The displacement memory can be derived \pj{from} the change of ${\sigma'}_s^0$ at early time $u_i'$ and late time $u_f'$ and is given by
\begin{multline}\label{dmemory}
\Delta {\sigma'}_s^0\mid_{u_i'}^{u_f'}=P_sP'(u'_f,z',\bz')^{-\frac32}\bP'(u'_f,z',\bz')^{\frac12}\sigma'^0(u'_f,z',\bz') \\
-P_sP'(u'_i,z',\bz')^{-\frac32}\bP'(u'_i,z',\bz')^{\frac12}\sigma'^0(u'_i,z',\bz')
+ P_s\p_{\bz'}^2\left[\int^{u'_f}_{u'_i} dv \;\sqrt{P'\bP'}\right].
\end{multline}

We would like to comment more, from the geometric point of view, on the case of \pj{an} arbitrary 2 surface. From the geodesic equation
\be
\n_n n=-(\gamma+\bar \gamma)n_\mu + \bar\nu \bm_\mu + \nu m_\mu,
\ee
we find that $n$ will not be tangent to a null geodesic on null infinity when $P$ is $u$-dependent. ${\mu}^0$ and ${\lambda}^0$ are the asymptotic expansion and shear of \pu{$n$ respectively \cite{Newman:1961qr}}. In the unit 2 sphere case, $n$ is tangent to null geodesic\pj{s} with affine parameter $u$. The Weyl tensor $\Psi_{s3}^0$ and $\Psi_{s4}^0$ are completely determined by the asymptotic shear of the null geodesic congruence. We call them \pj{the} \textit{news} functions as they indicate the existence of gravitational waves. However, in the case of arbitrary 2 surface, the asymptotic shear ${\lambda}^0$ will not only be controlled by gravitational waves, but will be also affected by the reference system. Nevertheless, we can define
\be
{\lambda}_s^0=P_s^2\left(\frac{\lambda^0}{\bP^2} + \frac{ \p_z^2\sqrt{P\bP}}{\sqrt{P\bP}}\right),
\ee
so that ${\lambda}_s^0$ measures only the gravitational wave contribution, while the second piece in the parentheses on the right hand side is purely the reference system effect.

\pu{The special choice of boundary topology in the standard treatment of the memory effect represents physical space-times which \pj{contain} isolated systems, \pj{with} no geometrical or topological information from outside world \cite{Geroch:1977jn}. An arbitrary 2 surface boundary is certainly compatible with the condition that no geometrical information \pj{is} coming from \pj{the} outside world. However, topological information can not be avoided, for instance \pj{a} null geodesic on \pj{null} infinity may not be complete. Hence the Weyl transformations are usually singular in such case. Another interesting transformation that \pj{involves} singularities are the so-called super-rotations \cite{Barnich:2009se,Barnich:2010eb,Barnich:2011ct}. The physical status of finite super-rotations \pj{and the transition between such states is} related to the breaking of a cosmic string via quantum black hole pair nucleation \cite{Strominger:2016wns}. It is definitely of interest to investigate the physical consequence of super-rotations in memory effects elsewhere.}

\section{Robinson-Trautman waves}

In this short section, we are ready to clarify the puzzle about the memory effect of the Robinson-Trautman waves via the generalized displacement memory formula developed in \pj{the} previous section. \pj{The} Robinson-Trautman metric was originally derived in \cite{Robinson:1960zzb} to demonstrate a very simple kind of spherical radiation. Adapted to our notation, the metric is\footnote{To compare with the solution in \cite{Newman:2009}, $P$ should \pu{be multiplied by a factor of $\frac12$.}}
\be
ds^2=2\left(-r\p_u \ln P + P^2\p_z\p_{\bz}\ln P + \frac{\Psi_2^0}{r} \right)du^2+2dudr-2\frac{r^2}{P^2}dzd\bz,
\ee
where $\Psi_2^0$ is a real constant and $P$ is a real arbitrary function of $(u,z,\bz)$ satisfying
\be\label{RTequation}
3\Psi^0_2\p_u P + P^3\p_{\bz}^2 P \p_z^2 P - P^4\p_z^2\p_{\bz}^2 P=0.
\ee
In \pj{the} NP formalism, the solution is given by
\begin{align}\label{RT}
&\Psi_0=\Psi_1=\sigma=\lambda=\tau=X^A=\omega=0,\nn\\
&\Psi_2=\frac{\Psi_2^0}{r^3},\;\;\;\;\mu_0=-P^2 \p\xbar\p \ln P\nn\\
&\Psi_3=\frac{P\p_z \mu^0}{r^2},\;\;\;\;\Psi_4=\frac{-\p_z(P^2\p_z\p_u\ln P)}{r}-\frac{P^2\p_z\p_{\bz}\mu^0}{r^2},\nn\\
&\rho=-\frac{1}{r},\;\;\;\;\alpha=\frac{\p_z P}{2r},\;\;\;\;\beta=-\frac{\pb P}{2r},\;\;\mu=\frac{\mu_0}{r} - \frac{\Psi^0_2}{r^2},\\
&\gamma=-\half \p_u \ln P-\frac{\Psi^0_2}{2r^2},\;\;\nu=-P\p_z\p_u\ln P-\frac{P\p_z \mu^0}{r},\nn\\
&U=r \p_u \ln P + \mu^0-\frac{\Psi^0_2}{r},\;\;L^z=0,\;\;L^{\bz}=\frac{P}{r}.\nn
\end{align}
As explained in \pj{the} previous section, $n$ is not tangent to a null geodesic on \pj{null} infinity in this case. Both gravitational radiation and the effect of \pj{the} reference system will contribute to the asymptotic shear of the null congruence ${\lambda}^0$ that $n$ is tangent to. Their contributions happen to cancel, namely ${\lambda}^0=0$. The standard displacement memory formula does not apply in this situation. To eliminate the reference effect, one needs to use a ``good'' reference system with time coordinate $\tilde{u}=\int^u_0 dv \;\frac{P}{P_s}$. However, according to \eqref{dmemory}, the displacement memory effect from gravitational radiation can be obtained directly in the original coordinates. It is just
\be
P_s\p_{\bz}^2\left(\int^{u_f}_{u_i} dv \;P\right).
\ee
To the best of our knowledge, non-trivial explicit solutions of equation \eqref{RTequation} \pj{rarely} exist. One may need to turn to numerical methods to find the exact value of displacement memory.

\section{Spin memory effect}
Recently, Pasterski, Strominger, and Zhiboedov discovered a new type of gravitational memory, the so-called spin memory effect \cite{Pasterski:2015tva}. They proposed that the observational effect of spin memory is the relative time delay of different light rays at very large radial distance $r_0$. In this section, we will provide a new observational effect by looking at time-like geodesics with affine parameter very close to null infinity.

The observer will be constrained \pj{to} a fixed radial distance $r_0$ which is very far from the gravitational source, \textit{e.g. on the earth}. The $r=r_0$ hypersurface is time-like, its induced metric can be derived easily from the most general NU solutions in Appendix \ref{NU solution}. Up to relevant orders, the induced metric is given by
\begin{multline}\label{metric}
ds^2=\left[1+\frac{\Psi_2^0 + \xbar \Psi_2^0}{r}-\frac{\xbar\eth\Psi_1^0 + \eth\xbar \Psi_1^0}{r^2} + O(r^{-3})\right]du^2 \\
- 2 \left[\frac{\eth \xbar\sigma^0}{P_s} - \frac{2 \xbar\Psi_1^0}{3P_s r} + O(r^{-2})\right] du dz
 - 2 \left[\frac{\xbar\eth \sigma^0}{P_s} - \frac{2 \Psi_1^0}{3P_s r} + O(r^{-2})\right] du d\bz \\
 - \left[2\frac{\xbar\sigma^0r}{P^2_s} - \frac{\xbar\Psi_0^0}{3P_s r} + O(r^{-2})\right]dz^2- \left[2\frac{\sigma^0r}{P^2_s} - \frac{\Psi_0^0}{3P_s r} + O(r^{-2})\right]d\bz^2\\
 -2\left[\frac{r^2}{P^2_s} + \frac{\sigma^0\xbar\sigma^0}{P^2_s}  + O(r^{-2})\right]dzd\bz.
\end{multline}
We will work in the unit 2 sphere case. \pu{An arbitrary 2 surface can be mapped onto \pj{the} unit 2 sphere by a Weyl transformation as discussed} in previous section (see more details in \cite{Barnich:2016lyg}).

Free falling observers on this hypersurface will travel along time-like geodesics. Supposing that \pj{the} vector $V$ is tangent to a time-like geodesic, it should satisfy the geodesic equation
\be
\xbar\n_V V=0,
\ee
where $\xbar\n$ is the covariant derivative on this 3 dimensional hypersurface. Actually, $V$ is induced from a 4 dimensional vector $\tilde V$. When $r\rightarrow\infty$, a 4 dimensional time-like vector $\tilde V$ will either vanish or \pu{be} proportional to the null basis $n$ (in \pj{the} Newman-Penrose formalism) which is the generator of null infinity, because the light-cone will be squashed to a line on null infinity. Hence $V$ should have the following asymptotic behavior:
\be
V^u=1 + \sum\limits_{a=1}^\infty\frac{V^u_a}{r^a},\;\;\;\;V^z=\sum\limits_{a=2}^\infty\frac{V^z_a}{r^{a}}.
\ee
Then we need to solve the geodesic equation order by order. The solution is (up to relevant order):
\be
\begin{split}
&V^u_1=-\frac{\Psi_2^0 + \xbar \Psi_2^0}{2}+V_{1I}^u(z,\bz)\\
&V^u_2=\frac16(\xbar\eth\Psi_1^0 + \eth \xbar\Psi_1^0) - \eth\xbar\sigma^0\xbar\eth\sigma^0 + \frac38(\Psi_2^0 + \xbar \Psi_2^0)^2 - \frac12V_{1I}^u(\Psi_2^0 + \xbar \Psi_2^0)+V_{2I}^u(z,\bz)\\
&V^z_2=-P_s \xbar\eth\sigma^0+V_{2I}^z(z,\bz) ,\\
&V^z_3=P_s\left[2\eth\xbar\sigma^0\sigma^0 + \frac23\Psi_1^0 + \frac12\xbar\eth\sigma^0 (\Psi_2^0 + \xbar \Psi_2^0) \right]-P_s\int\;dv\; \frac{\eth(\Psi_2^0 + \xbar \Psi_2^0)}{2} \\
&\hspace{7.5cm}- P_s\xbar\eth\sigma^0 V_{1I}^u - 2\sigma^0 V_{2I}^{\bz} + V_{3I}^z(z,\bz),\nn
\end{split}
\ee
where $V_{1I}^u,\;\;V_{2I}^u,\;\;V_{2I}^z,\;\;V_{3I}^z$ are integration constants that indicate the initial velocity of the observer. We will now set all of them to be zero as we require the observer is static initially.
At $r_0^{-2}$ order, $V$ has angular components due to the presence of gravitational waves characterized by $\sigma^0$. In other words, gravitational radiation force\pj{s} free falling time-like particles \pj{to rotate}. Since $V$ is time-like, the infinitesimal change of the proper time of the observer can be derived from the co-vector. It is just:
\be\begin{split}
d\chi=&du + \frac{1}{2r_0}\left((\Psi^0_2+\xbar\Psi^0_2) du + \frac{\int \;\xbar\eth (\Psi^0_2+\xbar\Psi^0_2) \;dv}{P_s} dz + \frac{ \int \;\eth (\Psi^0_2+\xbar\Psi^0_2) \;dv}{P_s} d\bz \right)+O(r_0^{-2})\\
=&d\left(u + \frac{1}{2r_0} \int\;(\Psi^0_2+\xbar\Psi^0_2) \;dv\right)+O(r_0^{-2}),
\end{split}\ee
where $\chi$ is the proper time. Defining
\be
\cM=\frac{1}{2}\int\;(\Psi^0_2+\xbar\Psi^0_2) \;dv,
\ee
between two space-time points $(u_i,z_i,{\bz}_i)$ and $(u_f,z_f,{\bz}_f)$ on this geodesic, the change \pj{in} proper time is
\be
\Delta \chi=\Delta u + \frac{1}{r_0} \Delta \cM+O(r_0^{-2}).
\ee
Clearly, the observer receives \pj{a} time delay at order $\frac1r$. We want to emphasize that $\Delta \cM$ is \pj{angle} dependent. Thus, both the Bondi mass aspect of \pj{a} massive object and gravitational radiation contribute to the time delay. $\Delta \cM$ is constrained by the time evolution equation
\be
\p_u\Psi^0_1=\eth\Psi^0_2 - 2\sigma^0 \p_u \eth{\xbar\sigma^0},
\ee
and it is completely fixed by the change of the angular momentum aspect as\footnote{See also \cite{Pasterski:2015tva} for the relation of the angular momentum flux \pj{to} the spin memory.}
\be\label{memory}
\xbar\eth\eth \Delta\cM =\Delta(\frac12\xbar\eth\Psi_1^0 + \frac12\eth\xbar\Psi_1^0 + \xbar\eth\sigma^0 \eth\xbar\sigma^0) + \int\;(\sigma^0\xbar\eth\dot\sigma^0 + \xbar\sigma^0\eth\xbar\eth\dot\sigma^0)\;dv,
\ee
up to a real constant. This is very similar to the displacement memory \cite{Christodoulou:1991cr,Frauendiener} in the sense that it includes the linear piece
\be
\Delta(\frac12\xbar\eth\Psi_1^0 + \frac12\eth\xbar\Psi_1^0 + \xbar\eth\sigma^0 \eth\xbar\sigma^0),
\ee
and the non-linear piece
\be
\int\;(\sigma^0\xbar\eth\dot\sigma^0 + \xbar\sigma^0\eth\xbar\eth\dot\sigma^0)\;dv.
\ee
The effect of the integration constant in \eqref{memory} can be eliminated by choosing a ring of freely falling observers who are initially static and synchronized\footnote{\pu{It is meaningful to point out that displacement memory and spin memory happen at the same time when  $\Delta \cM$ is \pj{angle} dependent. There is only displacement memory no spin memory when $\Delta \cM$ has no angular dependence. In such case, the changes \pj{in} proper time of \pj{a} ring of freely falling observers are the same. In this sense, we say displacement memory does not contribute to a relative time delay for a ring of free falling observers (see also Appendix A of \cite{Strominger:2014pwa}).}}. The difference \pj{in} the proper time of every observer on this ring at \pj{a} later time can \textit{memorize} the waveform of the gravitational waves. This is another observational effect of the spin memory.

Since we have $V=n$ when $r\rightarrow\infty$, the leading piece of the shear of the time-like geodesic congruence is just $\frac{\lambda^0}{r}$. The ring of free falling observers will observe two memory effects: the displacement memory that will squash and stretch the ring and the time delay that will cause the difference of the proper time of nearby observers.

\section{Conjugate points}

To measure the time delay effect, one may expect an even simpler ideal experiment if there are conjugate points on the time-like geodesics, which is possible in curved space-time. Two free falling observers are launched with certain initial velocities at the same point. Then they will meet again at the conjugate point where they can compare their proper time. However, such \pj{an} experiment is extremely hard to arrange. Because conjugate points on a time-like geodesic are very very far from each other in the asymptotic region in our set-up, though a time-like geodesic does have conjugate points \cite{Hawking:1973uf} in such case\pj{s}. In order to show that, we will consider part of a time-like geodesic, namely the length of this part of the geodesic is much smaller than $r_0$. Then the $\frac1r$ expansion can \pj{be applied}. In the end, we will prove that there \pu{are} no conjugate points on this part of the geodesic.

A solution $T^c$ of the geodesic deviation equation
\be
\xbar\n_V(\xbar\n_V T^c)=-{R_{abd}}^c\,T^bV^aV^d.
\ee
is called \pj{a} Jacobi field on the geodesic that $V$ is tangent to. Two points $p$ and $q$ are conjugate points along the geodesic if there exists a non-zero Jacobi field $T^c$ along the geodesic that vanishes at $p$ and $q$ \cite{Hawking:1973uf}. According to the induced metric \eqref{metric}, ${R_{abd}}^c\,V^aV^d=O(r^{-1})$, so the leading piece of the Geodesic deviation equation is
\be
\p_u^2 T^a_0(u,z,\bz)=0,
\ee
where $T^a_0(u,z,\bz)$ is the leading term of $T^a$ in the $\frac1r$ expansion. Then $T^a_0$ is given by
\be
T^a_0=uT^a_{01}(z,\bz)+T^a_{02}(z,\bz).
\ee
Naively, one can find \pj{an infinite number of possible choices of} $T^a_0$ that allow two points $(u_i,z_i,\bz_i)$ and $(u_f,z_f,{\bz}_f)$ from the geodesic to be conjugate points, namely
\be\begin{split}
&u_i\,T^a_{01}(z_i,\bz_i)+T^a_{02}(z_i,\bz_i)=0,\\
&u_f\,T^a_{01}(z_f,\bz_f)+T^a_{02}(z_f,\bz_f)=0.
\end{split}\ee
However, the change of the angular coordinates on this geodesic is very tiny and proportional to $\frac1r$. Then the second equation $u_f\,T^a_{01}(z_f,\bz_f)+T^a_{02}(z_f,\bz_f)=0$ becomes
\be
u_f\,T^a_{01}(z_i,\bz_i)+T^a_{02}(z_i,\bz_i)+O(r^{-1})=0.
\ee
Hence the condition of having $(u_i,z_i,\bz_i)$ and $(u_f,z_f,{\bz}_f)$ be conjugate points is reduced to
\be\begin{split}
u_i\,T^a_{01}(z_i,\bz_i)+T^a_{02}(z_i,\bz_i)=0,\\
u_f\,T^a_{01}(z_i,\bz_i)+T^a_{02}(z_i,\bz_i)=0,
\end{split}\ee
but $u_i\neq u_f$. Obviously, there is no such solution.

Thus we can never find two points that are not very far from each other on one geodesic where $T^a=0$ at leading order. Conceptually, this is expected since geodesics in flat (Minkowski) space-times do not have conjugate points. Now we \pj{have just shown} that conjugate points \pu{are} not ``close'' to each other in the asymptotic region in asymptotically flat space-times though they do exist.

\section{Discussions}

In this work, we first derived the formula \pj{for the} displacement memory effect for the case with an arbitrary 2 surface boundary topology. Via a Weyl transformation, it can be mapped into the unit 2 sphere. Then the standard formula \pj{for the} displacement memory \pj{applies}. This leads us to a direct derivation of the displacement memory formula in the original form of the solutions. Secondly, we proposed a new observational effect of the spin memory. It is a time delay of time-like free falling observer\pj{s}.

The discovery of spin memory was originally inspired \pj{by} the connection between \pj{the} gravitational memory effect and Weinberg's soft graviton theorem. The displacement memory and spin memory correspond to the leading and sub-leading soft graviton theorem, respectively. The novel results in \cite{Cachazo:2014fwa} show that soft graviton theorems exist even at third order in the low-energy expansion. There are indeed some positive signs indicating a third gravitational memory effect. On the one hand, the two known memory formulas are completely controlled by the time evolution equations of the Weyl tensors:
\be
\p_u\Psi^0_2=- \p_u\eth^2 \xbar\sigma^0 - \sigma^0 \p_u^2\xbar\sigma^0,
\ee
and
\be
\p_u\Psi^0_1=\eth\Psi^0_2 - 2\sigma^0 \p_u \eth{\xbar\sigma^0},
\ee
in unit 2 sphere case. There is indeed a third time evolution equation
\be
\p_u\Psi^0_0=\eth\Psi^0_1+3\sigma^0\Psi^0_2.
\ee
The asymptotic shear $\sigma^0$ is constrained by the imaginary part of this equation through some tedious but not difficult calculations (it is more clear in \pj{the} linearized gravity case \cite{Conde:2016rom}). On the other hand, displacement and spin memories are related to the energy flux and the angular momentum flux through null infinity. Newman and Penrose \cite{Newman:1965ik,Newman:1968uj} discovered more gravitationally-conserved quantities that may \pj{account} for the possible third gravitational memory (see also recent relevant development\pj{s} \cite{Compere:2017wrj,Godazgar:2018vmm,Godazgar:2018qpq,Godazgar:2018n}).

We have shown the power of Weyl transformation\pj{s} with a precise example in this work. Actually, the action of the full BMS$_4$ group combined with Weyl transformation\pj{s} on the Newman-Unti solution space was given in \cite{Barnich:2016lyg}. It would be very meaningful to compute the transformation law of the BMS$_4$ current, especially the action of \pj{a} Weyl transformation elsewhere \cite{tocome}.

\section*{Acknowledgments}

The authors thank Glenn Barnich and Jun-Bao Wu for useful discussions and additionally Jun-Bao Wu again for critical comments on our computations. This work is supported in part by the China Postdoctoral Science Foundation (Grant No. 2017M620908), by the National Natural Science Foundation of China (Grant Nos. 11575286, 11731001, 11475179, and 11575202).

\appendix

\section{Newman-Unti solution in \pj{the} Newman-Penrose formalism}
\label{NU solution}
The Newman-Penrose formalism \cite{Newman:1961qr} is a tetrad formalism with a special choice of four null basis vectors $e_1=l,\;e_2=n,\;e_3=m,\;e_4=\bar{m}$, where $l$ and $n$ are real while $m$ and $\bm$ are complex conjugates of each other.
The null basis vectors satisfy orthogonality conditions $l\cdot m=l\cdot\bm=n\cdot m=n\cdot\bm=0$ and normalization conditions $l\cdot n=1,m\cdot\bm=-1$. The connection coefficients are called spin coefficients in \pj{the} NP formalism with special Greek symbols (we will follow the convention of \cite{Chandrasekhar}). The metric is constructed from
\be
g_{\mu\nu}=n_\mu l_\nu + l_\mu n_\nu - m_\mu {\bm}_\nu - m_\nu \bm_\mu.
\ee

In \pj{the} NP formalism, it is always possible to impose
\be\nn
\pi=\kappa=\epsilon=0,\,\,\rho=\bar\rho,\,\,\tau=\bar\alpha+\beta,
\ee
which means that $l$ is tangent to a null geodesic with affine parameter and the congruence of the null geodesic is hypersurface orthogonal, \textit{i.e.} $l$ will be proportional to the gradient of a scalar field. It is convenient to take this scalar field as coordinate $u=x^1$ and set the affine parameter as coordinate $r=x^2$. To satisfy \pj{the} orthogonality and normalization conditions, the basis vectors and the cotetrad must have the form
\begin{align}
&n=\frac{\p}{\p u} + U \frac{\p}{\p r} + X^A \frac{\p}{\p x^A},\;\;\;\;\;\;l=\frac{\p}{\p r},\;\;\;\;\;\;m=\omega\frac{\p}{\p r} + L^A \frac{\p}{\p x^A},\nn\\
&n=\big[-U-X^A(\xbar\omega L_A+\omega \bar L_A)\big]du + dr + (\omega\bar L_A+\xbar\omega L_A)dx^A,\nn\\
&l=du,\;\;\;\;\;\;m=-X^A L_A du + L_A dx^A,\nn
\end{align}
where $L_AL^A=0$, $L_A\bar L^A=-1$. The main condition of approaching flatness at infinity is $\Psi_0=\frac{\Psi_0^0}{r^5}+O(r^{-6})$. Newman and Unti \cite{Newman:1962cia} derived the most general solutions of \pj{the} NP system that preserve the conditions listed above. The asymptotic expansion of all components in stereographic coordinates $(z,\bz)$ \pj{is} given by:
\begin{align}
&\Psi_0=\frac{\Psi_0^0(u,z,\bz)}{r^5}+O(r^{-6}),\;\;\;\;\;\;\Psi_1=\frac{\Psi_1^0(u,z,\bz)}{r^4}-\frac{\xbar \eth \Psi_0^0}{r^5}+O(r^{-6}),\nn\\
&\Psi_2=\frac{\Psi_2^0(u,z,\bz)}{r^3}-\frac{\xbar \eth \Psi_1^0}{r^4}+O(r^{-5}),\;\;\;\;\;\;\Psi_3=\frac{\Psi_3^0}{r^2}-\frac{\xbar \eth\Psi_2^0}{r^3}+O(r^{-4}),\nn\\
&\Psi_4=\frac{\Psi_4^0}{r}-\frac{\xbar \eth \Psi_3^0}{r^2}+O(r^{-3}),\nn\\
&\rho=-\frac{1}{r}-\frac{\sigma^0\xbar\sigma^0}{r^3}+O(r^{-5}),\;\;\;\;\;\;\tau=-\frac{\Psi^0_1}{2r^3}+O(r^{-4}),\nn\\
&\sigma=\frac{\sigma^0(u,z,\bz)}{r^2}+ (\sigma^0\sigma^0\xbar\sigma^0 - \frac12 \Psi_0^0)r^{-4} + O(r^{-5}),\nn\\
&\alpha=\frac{\alpha^0}{r}+\frac{\xbar\sigma^0\xbar\alpha^0}{r^2}+\frac{\sigma^0\xbar\sigma^0\alpha^0}{r^3}+O(r^{-4}),\nn\\
&\beta=-\frac{\xbar\alpha^0}{r}-\frac{\sigma^0\alpha^0}{r^2}-\frac{\sigma^0\xbar\sigma^0\xbar\alpha^0+\half \Psi^0_1}{r^3}+O(r^{-4}),\nn\\
&\mu=\frac{\mu^0}{r} - \frac{\sigma^0\lambda^0+\Psi^0_2}{r^2}+ (\sigma^0\xbar\sigma^0 \mu^0 + \frac12 \xbar\eth \Psi_1^0)r^{-3} + O(r^{-4}),\nn\\
&\lambda=\frac{\lambda^0}{r}-\frac{\xbar\sigma^0 \mu^0}{r^2}+ (\sigma^0\xbar\sigma^0 \lambda^0 + \frac12 \xbar\sigma^0 \Psi_2^0)r^{-3} + O(r^{-4}),\nn\\
&\gamma=\gamma^0-\frac{\Psi^0_2}{2r^2}+\frac16(2\xbar\eth\Psi_1^0 + \alpha^0 \Psi_1^0- \xbar\alpha^0\xbar\Psi_1^0)r^{-3} + O(r^{-4}),\nn\\
&\nu=\nu^0-\frac{\Psi^0_3}{r}+\frac{\xbar \eth \Psi^0_2}{2r^2}+O(r^{-3}),\nn\\
&\nn\\
&X^z=\frac{\bP\Psi_1^0}{6r^3}+O(r^{-4}),\;\;\;\;\;\;\omega=\frac{\xbar \eth \sigma^0}{r}-\frac{\sigma^0\eth \xbar\sigma^0+\half \Psi^0_1}{r^2}+O(r^{-3}),\nn\\
&U=-r(\gamma^0+\xbar\gamma^0) + \mu^0-\frac{\Psi^0_2 + \xbar \Psi^0_2}{2r}+\frac16(\xbar\eth\Psi_1^0 + \eth\xbar\Psi_1^0)r^{-2} +  O(r^{-3}),\nn\\
&L^z=-\frac{\sigma^0 \bP(u,z,\bz)}{r^2} - \frac{\bP}{r^4}({\sigma^0}^2\xbar\sigma^0 - \frac16\Psi_0^0) +O(r^{-5}),\;\;\;\; L^{\bz}=\frac{P(u,z,\bz)}{r}+\frac{\sigma^0 \xbar\sigma^0 P}{r^3}+O(r^{-5}),\nn\\
&L_z=-\frac{r}{\bP}+O(r^{-3}),\;\;\;\;\;\; L_{\bz}=-\frac{\sigma^0}{P}+ \frac{\Psi_0^0}{6Pr^2} +O(r^{-3}),\nn
\end{align}
where
\begin{align}
&\alpha^0=\half \bP \p_z \ln P,\;\;\;\;\;\;\mu^0=-\half P \bP \p_z \p_{\bz} \ln P\bP,\nn\\
&\lambda^0= \p_u{\xbar\sigma^0} + \xbar \sigma^0 (3\gamma^0 - \xbar \gamma^0),\nn\\
&\gamma^0=-\half \p_u \ln \bP,\;\;\;\;\;\;\nu^0=\xbar \eth (\gamma^0+\xbar\gamma^0),\nn\\
&\Psi_2^0 - \xbar\Psi_2^0 = \xbar\eth^2\sigma^0 -\eth^2\xbar\sigma^0 + \xbar\sigma^0\xbar\lambda^0 -\sigma^0\lambda^0,\nn\\
&\Psi^0_3=\xbar \eth \mu^0 - \eth \lambda^0,\;\;\;\;\;\;\Psi^0_4=\xbar\eth \nu^0 - \p_u\lambda^0 - 4 \gamma^0 \lambda^0,\nn\\
&\nn\\
&\p_u\Psi^0_0 + (\gamma^0 + 5 \xbar \gamma^0)\Psi^0_0=\eth\Psi^0_1+3\sigma^0\Psi^0_2,\nn\\
&\p_u\Psi^0_1 + 2 (\gamma^0 + 2 \xbar \gamma^0)\Psi^0_1=\eth\Psi^0_2+2\sigma^0\Psi^0_3,\nn\\
&\p_u\Psi^0_2 + 3 (\gamma^0 + \xbar \gamma^0)\Psi^0_2=\eth\Psi^0_3 + \sigma^0\Psi^0_4,\nn\\
&\p_u\Psi^0_3 + 2 (2 \gamma^0 + \xbar \gamma^0)\Psi^0_3=\eth\Psi^0_4.\nn
\end{align}
The ``$\eth$'' operator is defined by
\begin{equation}\begin{split}
&\eth \eta^s=P\bP^{-s}\p_{\bz}(\bP^s \eta^s)=P\p_{\bz} \eta^s + 2 s\xbar\alpha^0 \eta^s,\\
&\xbar\eth \eta^s=\bP P^{s}\p_z(P^{-s} \eta^s)=\bP\p_z \eta^s -2 s \alpha^0 \eta^s,\nn
\end{split}\end{equation}
where $s$ is the spin weight of the field $\eta$. The spin weight\pj{s} of relevant fields are listed below in Table \ref{t1}.
\begin{table}[h]
\caption{Spin weights}\label{t1}
\begin{center}\begin{tabular}{|c|c|c|c|c|c|c|c|c|c|c|c|c|c|c|c|c|c}
\hline
& $\eth$ & $\p_u$ & $\gamma^0$ & $\nu^0$ & $\mu^0$ & $\sigma^0$ & $\lambda^0$  & $\Psi^0_4$ &  $\Psi^0_3$ & $\Psi^0_2$ & $\Psi^0_1$ & $\Psi_0^0$   \\
\hline
s & $1$& $0$& $0$& $-1$& $0$& $2$& $-2$  &
  $-2 $&  $-1$ & $0$ & $1$ & $2$    \\
\hline
\end{tabular}\end{center}\end{table}

\section{Weyl transformation of solutions}
\label{Weyl}
In \pj{the} Newman-Penrose formalism, a gauge transformation is a combination of a change of coordinates and a local Lorentz transformation which is described in the standard three classes of rotation\pj{s} \cite{Chandrasekhar}. The residual gauge transformation\pj{s} preserving the gauge condition and asymptotic behaviors of Newman-Unti solutions \pj{were} derived by Barnich and Troessaert \cite{Barnich:2016lyg} recently. A pure Weyl transformation is characterized by
\be
z'=z,\;\; u'=\int^u_0 dv e^{E_R},\;\; P'=P e^{-E_R}.\nn
\ee
The transformation law of the data that characterizes \pj{the} asymptotic solution is given by
\bea
&&\sigma'^0=e^{-E_R}\left[\sigma^0 + \eth(e^{-E_R}\eth u') - (e^{-E_R}\eth u')(\p_u + \xbar \gamma^0 - \gamma^0)(e^{-E_R}\eth u')  \right], \nn \\
&&\lambda'^0=e^{-2E_R}\left[\lambda^0 + (\p_u + 3\gamma^0 - \xbar \gamma^0)[\xbar\eth (e^{-E_R}\xbar\eth u') - (e^{-E_R}\xbar\eth u')(\p_u + \gamma^0 - \xbar \gamma^0)(e^{-E_R}\xbar\eth u')]\right], \nn \\
&&\Psi'^0_4=e^{-3E_R } \left[\Psi_4^0 \right], \nn \\
&&\Psi'^0_3=e^{-3E_R}\left[\Psi_3^0 - e^{-E_R}\eth u' \Psi_4^0 \right], \nn\\
&&\Psi'^0_2=e^{-3E_R}\left[\Psi_2^0 - 2 e^{-E_R}\eth u' \Psi_3^0 + (e^{-E_R}\eth u')^2 \Psi_4^0 \right], \nn \\
&&\Psi'^0_1=e^{-3E_R}\left[\Psi_1^0 - 3 e^{-E_R}\eth u' \Psi_2^0 +3 (e^{-E_R}\eth u')^2 \Psi_3^0 - (e^{-E_R}\eth u')^3 \Psi_4^0 \right], \nn \\
&&\Psi'^0_0=e^{-3E_R}\left[\Psi_0^0 - 4 e^{-E_R}\eth u' \Psi_1^0 + 6 (e^{-E_R}\eth u')^2 \Psi_2^0 - 4 (e^{-E_R}\eth u')^3 \Psi_3^0 + (e^{-E_R}\eth u')^4 \Psi_4^0 \right]. \nn
\eea

\bibliography{ref}

\bibliographystyle{utphys}

\end{document}